# Insulators at Fractional Fillings in Twisted Bilayer Graphene Partially Aligned to Hexagonal Boron Nitride

Dillon Wong[1,*], Kevin P. Nuckolls[1,*], Myungchul Oh[1,*], Ryan L. Lee[1], Kenji Watanabe[2], Takashi Taniguchi[3], Ali Yazdani[1,‡]

[1] *Joseph Henry Laboratories and Department of Physics, Princeton University, Princeton, NJ 08544, USA*
[2] *Research Center for Functional Materials, National Institute for Materials Science, 1-1 Namiki, Tsukuba 305-0044, Japan*
[3] *International Center for Materials Nanoarchitectonics, National Institute for Materials Science, 1-1 Namiki, Tsukuba 305-0044, Japan*

\* These authors contributed equally to this work.
‡ Corresponding author email: yazdani@princeton.edu

**ABSTRACT**

**At partial fillings of its flat electronic bands, magic-angle twisted bilayer graphene (MATBG) hosts a rich variety of competing correlated phases that show sample to sample variations. Divergent phase diagrams in MATBG are often attributed to the sublattice polarization energy scale, tuned by the degree of alignment of the hexagonal boron nitride (hBN) substrates typically used in van der Waals devices. Unaligned MATBG exhibits unconventional superconductivity and correlated insulating phases, while nearly perfectly aligned MATBG/hBN exhibits zero-field Chern insulating phases and lacks superconductivity. Here we use scanning tunneling microscopy and spectroscopy (STM/STS) to observe gapped phases at partial fillings of the flat bands of MATBG in a new intermediate regime of sublattice polarization, observed when MATBG is only partially aligned ($\theta_{Gr\text{-}hBN} \approx 1.65°$) to the underlying hBN substrate. Under this condition, MATBG hosts not only phenomena that naturally interpolate between the two sublattice potential limits, but also unexpected gapped phases absent in either of these limits. At charge neutrality, we observe an insulating phase with a small energy gap ($\Delta <$ 5 meV) likely related to weak sublattice symmetry breaking from the hBN substrate. In**

**addition, we observe new gapped phases near fractional fillings v = ±⅓ and v = ±⅙, which have not been previously observed in MATBG. Importantly, energy-resolved STS unambiguously identifies these fractional filling states to be of *single-particle origin*, possibly a result of the super-superlattice formed by two moiré superlattices. Our observations emphasize the power of STS in distinguishing single-particle gapped phases from many-body gapped phases in situations that could be easily confused in electrical transport measurements, and demonstrates the use of substrate engineering for modifying the electronic structure of a moiré flat-band material.**

Van der Waals moiré materials display a variety of correlated insulating phases at densities corresponding to integer multiples of the moiré-Brillouin-zone area (integer fillings v)[1–8]. Such integer-v correlated insulators are widely observed in a multitude of moiré material devices, and they are often the first signatures of the strong electronic correlations that are ubiquitous in flat-band moiré systems[9,10]. Less common are correlated insulating phases at densities corresponding to non-integer multiples of the moiré-Brillouin-zone area (non-integer v). Such phases are occasionally reported in twisted graphene systems[11–15] and transition metal dichalcogenide (TMD) hetero/homobilayers[16–19]. Non-integer filling phases are conjectured to involve broken space group symmetries (i.e. broken translational/rotational symmetry, where the total density at one moiré site differs from the density at a neighboring moiré site) with long-range spatial order that can only be stabilized under the highest levels of material quality and homogeneity[20–22]. However, these non-integer-v phases are primarily detected through electrical resistivity, optical reflectance, and compressibility measurements, and the natures of their ground states are only inferred from the simplicity of their rational filling factors. To date, only $WSe_2/WS_2$ and twisted $WS_2$ have been confirmed to host generalized Wigner crystal states at v = ⅓ with direct spatial imaging[23,24].

Here we use scanning tunneling microscopy and spectroscopy (STM/STS) to characterize magic-angle twisted bilayer graphene (MATBG) that is partially aligned to its hexagonal boron nitride (hBN) substrate ($\theta_{Gr\text{-}hBN} \approx 1.65°$). We report a sequence of non-integer-v phases at simple fractions v = ±⅓ and v = ±⅙ in an MATBG device, seen as strong suppressions of the zero-bias conductance at these fillings. Although they may appear to be previously unobserved correlated insulating states, tunneling conductance $dI/dV(V_s, V_g)$ as a function of sample bias $V_s$ and gate voltage $V_g$ reveals that these non-integer-v gapped phases are *not* correlated insulators, but rather are most likely of single-particle origin. Scanning tunneling microscopy (STM) topographic images show a significant moiré pattern between graphene and hBN, which strongly modifies the Gr-Gr moiré pattern and suggests that the $v = \pm\frac{1}{3}$ and $v = \pm\frac{1}{6}$ phases may be associated with a very long wavelength super-superlattice formed from the graphene-graphene (Gr-Gr) and the graphene-hBN (Gr-hBN) moiré patterns[25–28]. In addition, we observe an insulating gap at the charge neutrality point (CNP; $v = 0$) that is likely due to sublattice symmetry breaking from the hBN substrate, as well as an extended, spatially inhomogeneous pseudogap-like feature. Our results highlight the importance of spatial imaging and spectroscopy in elucidating the origin of gapped states in moiré materials.

Our measurements were acquired on a homebuilt dilution-refrigerator STM[29] using a W tip calibrated against the surface state of Cu(111), with scanning tunneling spectroscopy (STS) obtained using a standard lock-in technique. The MATBG device was prepared using a method similar to that described elsewhere[30]. MATBG rests on an hBN/SiO$_2$/Si substrate, and $V_g$ is applied to the Si layer to tune the electron density in the MATBG. The MATBG is electrically connected to prepatterned Au/Ti electrodes biased at $V_s$.

Fig. 1(a) shows a topographic image of MATBG/hBN. The triangular lattice of dark sites with periodicity $\lambda_{Gr-hBN}$ = 7.4 nm corresponds to the moiré pattern between the graphene and hBN atomic lattices, while the bright regions (e.g. labeled A and B) are the AA sites of the MATBG moiré pattern (with period $\lambda_{Gr-Gr}$ = 15.2 nm, $\theta_{Gr-Gr} \approx 0.93°$). The Fourier transform of this image (inset) shows clear peaks for Gr-Gr and Gr-hBN reciprocal lattice wave vectors, which are rotated from one another by approximately 30° [31,32]. dI/dV($V_s$, $V_g$) on spot A (Fig. 1(b)) shows a pair of flat bands that are pinned to the Fermi energy ($E_F$, $V_s$ = 0 mV) between Gr-Gr filling factors $v$ = +4 and $v$ = -4. Gap-like features appear at $E_F$ for 1 < $v$ < 3 and for -3.5 < $v$ < -2, possibly related to the pseudogap observed in MATBG unaligned with hBN[33,34], which we will discuss later.

Focusing on the CNP, Figs. 1(c-d) show a small gap ($\Delta$ < 5 meV) convolved with tip-induced charging effects that confirm the insulating nature of this state. This gap parallels STS on perfectly aligned MATBG, which identifies a large-energy gapped spectrum ($\Delta$ > 20 meV) near the CNP, again convolved with tip-induced charging effects[33]. When the underlying hBN substrate is aligned to MATBG, it breaks graphene's sublattice symmetry, gaps the Dirac nodes in the spectrum of MATBG, and produces the insulating gap observed at this filling[35,36]. A natural interpretation of the smaller gap observed here is that the partial alignment of the underlying hBN substrate breaks the sublattice symmetry of MATBG, but to a lesser degree than in perfectly aligned devices. Indeed, similar phenomena have been observed in transport experiments on graphene/hBN, where the extracted size of the sublattice symmetry-broken gap was observed to continuously increase with decreasing rotational mismatch between graphene and hBN[27,37,38]. The gap at the CNP contrasts with STS experiments on unaligned MATBG devices, which instead identify a small, but finite local density of states (LDOS), indicative of the commonly observed gapless semimetallic state expected in the presence of Dirac nodes[30,39].

In addition, zero-bias conductance dI/dV($V_s$ = 0 V, $V_g$) in Figs. 1(e-f) shows deep suppressions of the LDOS at ν = ±⅓ (yellow shaded bars) and weak suppressions of the LDOS at ν = ±⅙ (blue shaded bars). In the absence of bias-resolved spectroscopy, these zero-bias conductance suppressions would be highly suggestive of gapped insulating states derived from electron-electron interaction effects, a hypothesis that hinges on the simplicity of their rational filling factors and their presence at partial fillings of a flat electronic band. However, bias-resolved STS offers crucial information that contradicts this hypothesis. dI/dV($V_s$, $V_g$) in Figs. 1(c-d) shows that the ν = ±⅓ insulating gaps are observed away from the Fermi level as ν is tuned away from ±⅓, indicating that they are single-particle gapped phases that are rigidly tuned to the Fermi level via the gate voltage (also true for ν = ±⅙). This is in strong contrast to STS measurements of correlation-driven insulating phases in MATBG, which are observed to spontaneously appear at the Fermi level, often in place of an expected peak in the single-particle LDOS. An example of this is observed in the spectroscopic properties of the correlated insulator at ν = -2 in unaligned devices, where an insulating gap is measured to open and close only around $E_F$[33]. Thus, an alternate hypothesis is necessary for explaining these single-particle insulators at fractional fillings.

A plausible explanation for the non-integer-ν gaps stems from the similarity of two observed moiré wavelengths, $\lambda_{Gr-Gr}$ and $2\lambda_{Gr-hBN}$, which can produce a super-superlattice from the interference of the Gr-Gr and Gr-hBN moiré patterns. A simulated image in Fig. 2(a) that uses the length scales measured from the Fig. 1(a) topograph shows such a super-superlattice. Here, we plot the "super-superlattice moiré function"

$$F(\mathbf{r}) = \left(3 - \sum_{i=1}^{3} \cos R_i G_{Gr-hBN} \mathbf{r}\right)\left(3 + \sum_{i=1}^{3} \cos R_i G_{Gr-Gr} \mathbf{r}\right)$$

where $G_{Gr-hBN}$ ($G_{Gr-Gr}$) are the primitive reciprocal lattice vectors of the Gr-hBN (Gr-Gr) moiré pattern and $R_i = C_3^i$ are the 120° rotations. Visual inspection of the image reveals a large-scale approximate periodicity of the super-superlattice (marked by the yellow rhombus in Fig. 2(a) and by blue translation vectors in Fig. 2(c)) that is roughly 7 times longer than the Gr-hBN moiré pattern (or roughly 7/2 times larger than the Gr-Gr moiré pattern), denoted $\Lambda_1$. Accounting for the four-fold spin and valley degeneracies of the electronic states of MATBG, a triangular super-superlattice potential with length scale $\Lambda_1$ is expected to produce a partial gap at filling ν = ±4/($\Lambda_1$/$\lambda_{Gr-Gr}$)² ≈ ±4/(7/2)² = ±16/49 ≈ ±1/3, consistent with our observations of deep zero-bias conductance suppressions near fillings v = ±⅓. The "approximate translation symmetry function" $S(t) = 1/\int |F(r-t) - F(r)|^2 d^2r$ (plotted in Fig. 2(b)), which measures how well the super-superlattice moiré function $F(r)$ is preserved when translated by $t$, shows a strong local maximum at $t$ = (52.2 nm, 0), again consistent with the filling factor v = ±4/($\Lambda_1$/$\lambda_{Gr-Gr}$)² = ±0.34 ≈ ±1/3.

The origin of the v = ±⅙ spectroscopic gaps is less clear, but here we provide one possible hypothesis. Visual inspection of the simulated moiré super-superlattice indicates that unit cells of side length $\Lambda_1$ are well-approximated in many regions locally, but are insufficient for describing the entire super-superlattice periodicity. Instead, as shown in the tiling diagram in Fig. 2(c), the complete super-superlattice is characterized by two length scales, $\Lambda_1$ = 52.2 nm and $\Lambda_2$, = 77 nm, where the latter is observed to be roughly 10 times longer than the Gr-hBN moiré pattern (or roughly 5 times larger than the Gr-Gr moiré pattern). We find that $\Lambda_2$ also represents an approximate quasiperiodicity of the super-superlattice (marked by orange translation vectors in Fig. 2(c)), but to a lesser degree than the $\Lambda_1$ periodicity. A triangular super-superlattice potential with length scale $\Lambda_2$ is expected to produce a partial gap at filling v = ±4/($\Lambda_2$/$\lambda_{Gr-Gr}$)² ≈ ±4/(5)² = ±4/25 ≈ ±1/6, consistent with our observations of weaker zero-bias conductance suppressions near fillings v = ±⅙. In addition, the approximate translation symmetry function $S(t)$ (as well as the autocorrelation of the super-superlattice moiré function $F(r)$) shows a local maximum at $t$ = (0, 77 nm), which again corresponds to a filling factor v = ±4/($\Lambda_2$/$\lambda_{Gr-Gr}$)² = ±0.16 ≈ ±1/6. Despite the presence of these suppressions at v = ±1/6, we do not believe the suppressions found at v = ±1/3 are higher order gaps associated with $\Lambda_2$. Moreover, we do not observe any higher order gaps associated with $\Lambda_1$ and $\Lambda_2$, which may parallel observations of satellite features in graphene / hBN moiré superlattices[40]. We note, however, that this toy picture of the moiré super-superlattice may not be a completely accurate description of the system, especially since inhomogeneity in the twist angle over long length scales is reported in MATBG samples[41].

Further evidence that a super-superlattice potential influences the tunneling conductance is shown in Figs. 3(a,c), which show dI/dV($V_s$, $V_g$) for filling factors beyond full filling v = ±4. At these densities, we resolve a series of LDOS peaks in the dispersive remote bands that are unexpected from the band structure of MATBG. These peaks are more clearly seen in fixed-gate-voltage dI/dV($V_s$) spectra in Figs. 3(b,d) and are consistent with the emergence of the van Hove singularities of replica remote bands, which would result from Brillouin-zone folding in the presence of a super-superlattice potential modulation with a wavelength longer than the Gr-Gr moiré wavelength.

Finally, we turn to the density-extended gaps at $E_F$ seen between gate voltages 10 V < $V_g$ < 25 V and -25 V < $V_g$ < -10 V in spot A (Fig. 4(b)). These extended gaps are unlikely of single-particle origin, as they spontaneously appear at $E_F$ when the flat bands are partially filled, and they do not exist at energies away from $E_F$. In some ways, these gaps resemble the pseudogap features seen in MATBG unaligned with hBN, which are absent when MATBG is perfectly aligned with hBN[33]. However, the relationship between these gaps and the pseudogap is unclear at this time. Interestingly, the energy widths and gate voltage ranges of the extended gap-like features appear to vary dramatically with spatial position, as seen most strikingly when comparing spectroscopy obtained in locations A and B. In some areas (e.g. location B) the gap-like feature seems not to be present at all. We speculate that this spatial inhomogeneity could be related to the degree of local $C_2$-symmetry breaking, which varies among locations on the Gr-hBN moiré pattern[42–44], although understanding the relationship between structural and electronic properties in MATBG is challenging and requires a more careful study than was carried out here. Particularly, further studies are required to understand the spatial dependence of these gap-like features and to correlate their energy widths and gate voltage ranges to different locations on the superlattice.

Detailed band structure calculations and further measurements will be required to fully understand the nuanced effects of the sublattice potential on MATBG in partially aligned configurations with the underlying hBN substrate. Such measurements will be particularly important in understanding the relationship between superconductivity and $C_2$ rotational symmetry[45], which is critical to several open proposals for the mechanisms of this exotic superconductor[46–48]. Our spectroscopic measurements show that sublattice potential asymmetry should no longer be considered a matter of presence and absence, but rather a matter of the degree and the microscopic form of this crucial energy scale. Furthermore, our measurements highlight the importance of local spectroscopic probes in the field of moiré materials. STM/STS offers spatially and energy-resolved information that can vastly reduce the phase space of plausible explanations for new electronic phases observed in these materials.

**Acknowledgements**

We thank B. Andrei Bernevig for helpful discussions. This work was primarily supported by the Gordon and Betty Moore Foundation's EPiQS initiative grants GBMF9469 and ARO-MURI W911NF2120147. Other support for the experimental work was provided by DOE-BES grant DE-FG02-07ER46419, NSF-MRSEC through the Princeton Center for Complex Materials NSF-DMR- 2011750, NSF-DMR-1904442. K.W. and T.T. acknowledge support from the Elemental Strategy Initiative conducted by the MEXT, Japan, grant JPMXP0112101001, JSPS KAKENHI grant 19H05790 and JP20H00354.

# Figure 1

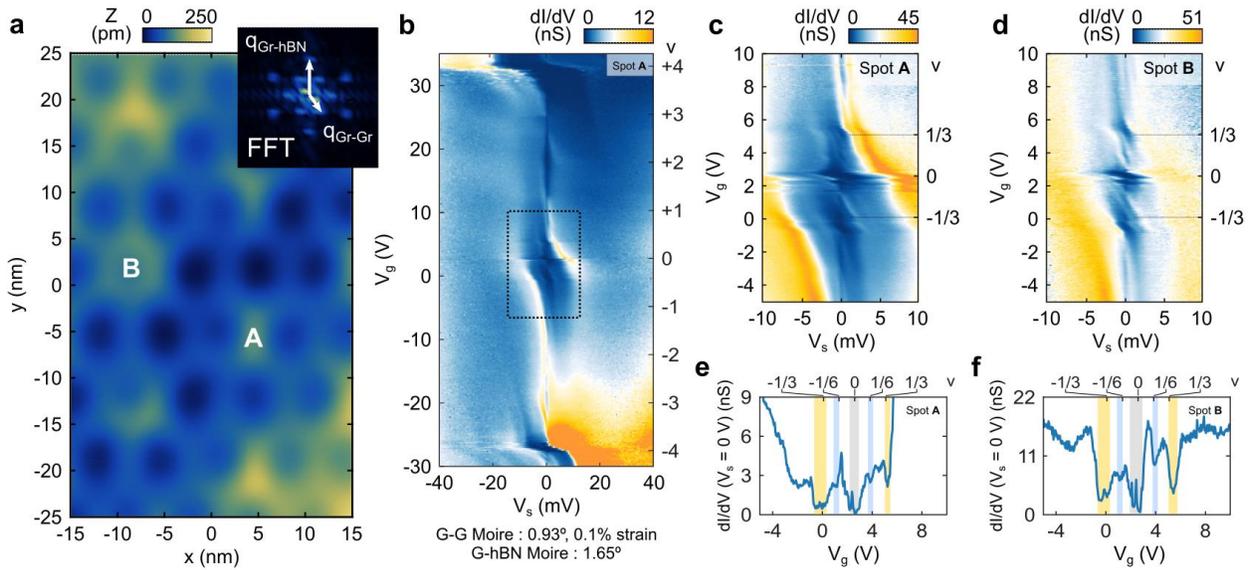

**Figure 1 | Single-Particle Gaps at Fractional Fillings in Partially Aligned Twisted Bilayer Graphene.** (a), STM topographic image of a magic-angle graphene moiré superlattice (bright regions) with a graphene-hBN moiré superlattice (dark regions). Inset: FFT of STM topograph, showing two moiré wavevectors, $Q_{Gr-hBN}$ and $Q_{Gr-Gr}$. (b), $dI/dV(V_s, V_g)$ measured at the center of the AA site labeled "A" in (a) at 200 mK and zero magnetic field. Dashed line box indicates the zoomed-in region shown in (c). (c), $dI/dV(V_s, V_g)$ measured at the center of the AA site labeled "A" in (a). (d), Same as (c), measured at the center of the AA site labeled "B" in (a). Zero-bias conductance $dI/dV(V_s = 0\ V, V_g)$ of data in (c). (f), Zero-bias conductance $dI/dV(V_s = 0\ V, V_g)$ of data in (d). Yellow shaded bars highlight deep conductance suppressions at $v = +\text{-}\tfrac{1}{3}$. Blue shaded bars highlight weak conductance suppressions at $v = +\text{-}\tfrac{1}{6}$. Tunneling parameters: (a) $V_s = -80$ mV, $I = 10$ pA, (b) $V_s = -80$ mV, $I = 500$ pA, $V_{A.C.} = 0.3$ mV at 381.7 Hz, (d) $V_s = -70$ mV, $I = 1$ nA, $V_{A.C.} = 0.15$ mV at 381.7 Hz.

# Figure 2

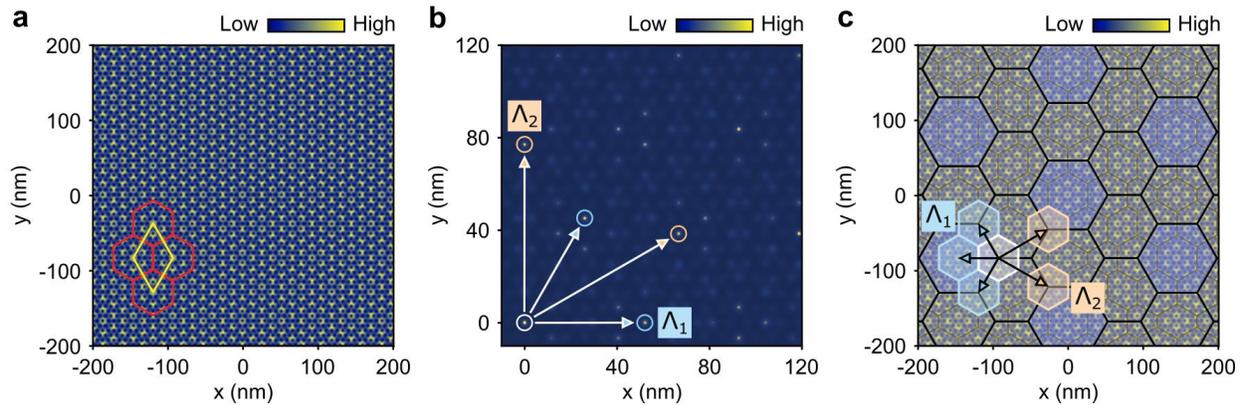

**Figure 2 | Schematic of Two Length Scales in Simulated Moiré Super-Superlattice.** (a), Simulated "super-superlattice moiré function" $F(r)$ using measured lengths and angles from STM topographic images. The red hexagons and the yellow rhombus depict the locally approximate super-superlattice unit cell. (b), Calculated "approximate translation symmetry function" $S(r)$, which shows a peak at a translation vector $t$ when $F(r)$ and $F(r-t)$ are most similar. Peaks in this function are observed when $t = \Lambda_1$ and $t = \Lambda_2$. The two unlabeled vectors are equivalent to vectors $\Lambda_1$ and $\Lambda_2$, related to the labeled vectors by a 60° rotation about the origin. (c), Annotated version of the plot in (a), illustrating the two approximate translation symmetry vectors, $\Lambda_1$ (blue vectors) and $\Lambda_2$ (orange vectors), and the two-length-scale tiling that covers the complete moiré super-superlattice using these two approximate translation vectors.

# Figure 3

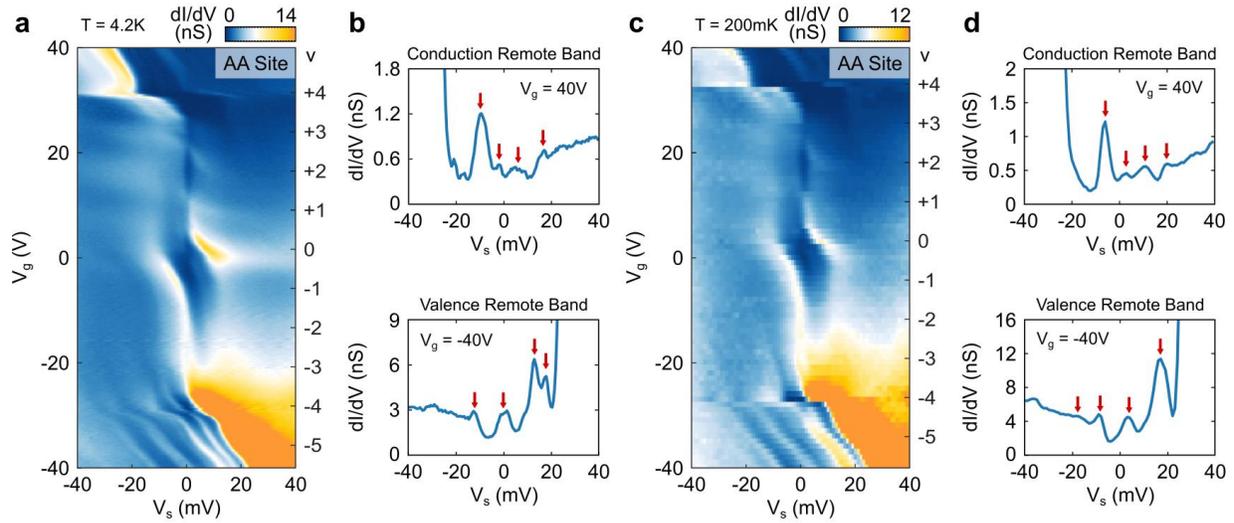

**Figure 3 | Evidence for Replica Remote Bands from a Moiré Super-Superlattice.** (a), dI/dV($V_s$, $V_g$) measured at the center of an AA site at T = 4.2 K. (b), dI/dV($V_s$) line cut spectra from (a) showing replica remote band peaks at fillings v > +4 (top) and v < -4 (bottom). (c), dI/dV($V_s$, $V_g$) measured at the center of an AA site at T = 200 mK. (d), dI/dV($V_s$) line cut spectra from (c) showing replica remote band peaks at fillings v > +4 (top) and v < -4 (bottom). Tunneling parameters: (a) $V_s$ = -80 mV, I = 300 pA, $V_{A.C.}$ = 0.5 mV at 381.7 Hz, (b) $V_s$ = -100 mV, I = 1 nA, $V_{A.C.}$ = 1 mV at 381.7 Hz.

# Figure 4

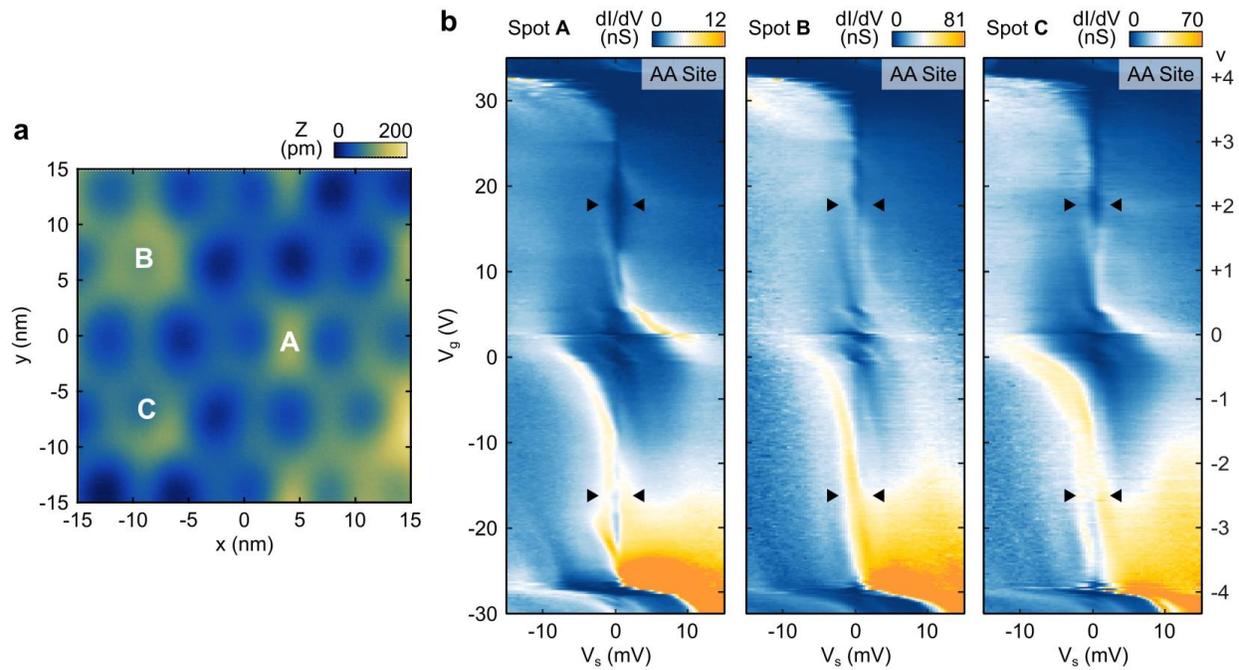

**Figure 4 | Signatures of Local Pseudogap-Like Features.** (a), STM topograph of MATBG partially aligned to hBN. (b), dI/dV($V_s$, $V_g$) measured at the center of AA sites at T = 200 mK, at locations A (left), B (middle), and C (right) as labeled in (a). Tunneling parameters: $V_s$ = -80 mV, I = 500 pA, $V_{A.C.}$ = 0.3 mV at 381.7 Hz for spot A; $V_s$ = -80 mV, I = 1 nA, $V_{A.C.}$ = 0.4 mV at 381.7 Hz for spot B; $V_s$ = -80 mV, I = 1 nA, $V_{A.C.}$ = 0.4 mV at 381.7 Hz for spot C.